\documentclass[]{iopart}
\usepackage{graphicx}
\usepackage{amssymb}
\usepackage[T1]{fontenc}
\usepackage[latin9]{inputenc}

\newcommand{\ket}[1]{|#1 \rangle}
\newcommand{\bra}[1]{\langle #1|}

\makeatletter

\begin{document}
%\title{Parallel interaction-free quantum imaging using spatial adiabatic passage: searching a quantum minefield.}
\title{Parallel interaction-free measurement using spatial adiabatic passage.}

\author{Charles Hill}
\address{Centre for Quantum Computation and Communication Technology, School of
Physics, The University of Melbourne, Melbourne, Victoria 3010,
Australia.}

\author{Andrew D.~Greentree}
\address{School of Physics, The University of Melbourne, Melbourne, Victoria 3010,
Australia.}

\author{Lloyd C.~L.~Hollenberg}
\address{Centre for Quantum Computation and Communication Technology, School of
Physics, The University of Melbourne, Melbourne, Victoria 3010,
Australia.}

\begin{abstract}
Interaction-free measurement is a surprising consequence of quantum
interference, where the presence of objects can be sensed without any
disturbance of the object being measured.  Here we show an extension
of interaction-free measurement using
techniques from spatial adiabatic passage, specifically multiple
reciever adiabatic passage.  Due to subtle properties of the adiabatic
passage, it is possible image an object without interaction between
the imaging photons and the sample.  The technique can be used on
multiple objects in parallel, and is entirely deterministic in the
adiabatic limit.  Unlike more conventional interaction-free measurement
schemes, this adiabatic process is driven by the symmetry of the
system, and not by more usual interference effects.  As such it
provides an interesting alternative quantum protocol which may be
applicable to photonic implementations of spatial adiabatic passage.
We also show that this scheme can be used to implement a
collision-free quantum routing protocol.
\end{abstract}


\maketitle

%%%%%%%%%%%%%%%%%%%%%%%%%%%%%%%%%%%%%%%%%%%%%%%%%%%%%%%%%%%%%%%%

\section{Introduction}  

One of the counter-intuitive effects of quantum
mechanics is that of interaction-free measurement (IFM).  Classically,
one can only obtain information about the location of an object by
interacting particles with it.  For example, by absorption or
deflection of the particles.  In every case there is an inevitable
interaction between the sensing particles, and the object being
sensed.  However, Dicke \cite{Dicke81}, and later Elitzur and Vaidman
\cite{EV93} showed that it is possible to perform an IFM, i.e. a
measurement where the sensing particles have never interacted with the
object, and yet provide definitive information about its existence.
To illustrate IFMs we summarise the canonical IFM: the `quantum bomb' problem.

Assume that we are searching for a bomb.  However this bomb is so
sensitive that if a single photon touches it, it will explode: an
undesirable outcome.  How can we design an optical sensing system that
detects the presence or absence of the bomb without setting it off?
Elitzur and Vaidman showed that this task can be achieved
non-deterministically.  They considered a Mach-Zehnder interferometer,
balanced so that when both arms are unobstructed, photons always exit
via the bright port.  However if a bomb is placed in one of the arms,
then there is a $50\%$ chance of detecting the photon at the dark
port, and hence detection at the dark port proves the presence of a
bomb without the photon having interacted with the bomb.  

The quantum bomb protocol is non-deterministic: the bomb is triggered
$50\%$ of the time, and detection of a photon at the bright port does
not provide information about the presence or absence of the bomb.
Nonetheless this is an important protocol which performs a task not
possible classically.  The success probability of this protocol can be
asymptotically increased to unity \cite{KWH+95}.  It is also possible
to perform multiple imaging and this was used to demonstrate
high-resolution images \cite{WMN+98}.  Variants of IFM lead to the
possibility of counter-factual quantum computation, where the result
of a quantum algorithm is determined without the qubits actually
performing the algorithm \cite{MJ01}.

\begin{figure}
\begin{centering}
\includegraphics[width=0.6\columnwidth]{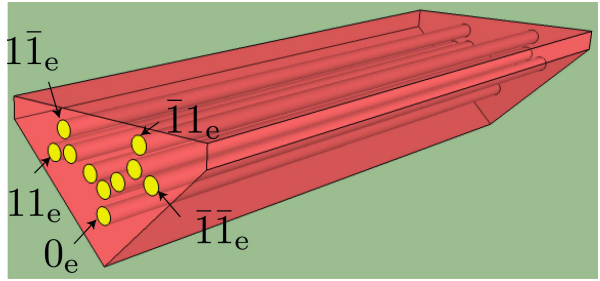}
\par\end{centering}
\caption{Multi-waveguide concept diagram for a four-leaf MRAP
  device, such as can be realised using direct-write
  technology. Waveguides are represented as circular tubes through the
  device.  In this device, the adiabatic protocol coherently sends photons from the input node, $\ket{0_e}$ to an equally weighted superposition $(1/2)(\ket{11_e} + \ket{1\bar{1}_e} + \ket{\bar{1}1_e} + \ket{\bar{1}\bar{1}})$, as indicated.\label{fig:Waveguide}}
\end{figure}

All of the above protocols have at their heart interferometry.  As
such they are sensitive to issues such as alignment and/or timing.  We
are considering an alternative approach to IFM based on adiabatic
passage.  As such, instead of relying on interferometry to perform the
measurement, we use the symmetry of the problem, in particular the
composition of the null space of the solution to the adiabatic
network, as will be described below.  This gives a fundamentally
different approach to the task of IFM, which gives rise quite
naturally to parallel search with robustness and deterministic
sensing.  Returning to the quantum bomb analogy, we can think of our
scheme as allowing the imaging of a quantum minefield, with multiple
quantum bombs distributed at unknown spatial locations.  The adiabatic
network can best be achieved using an integrated multi-waveguide
approach, leveraging the advances direct-write lithography
\cite{MPM+09} for quantum photonics \cite{Politi02052008}, and as such
can be seen as an extension of the demonstrations of waveguide
Coherent Tunneling Adiabatic Passage (CTAP) by Longhi and co-workers
\cite{Lon07,LDVOL07,VOF+08}.  A concept diagram of our envisaged device is shown in Fig.~\ref{fig:Waveguide}.

Techniques for adiabatic passage are used to evolve quantum states so
that the system remains in an instantaneous eigenstate, but the
Hamiltonian is varied as a function of time.  The requirement for
adiabaticity implies that the Hamiltonian must be changed sufficiently
slowly so that population is unable to leak between the eigenstates.
Such techniques are well known, especially in atomic and molecular
systems \cite{VHSB01}.  Advances in the construction of quantum
systems have allowed new perspectives in adiabatic passage, in
particular the opportunity the to move particles adiabatically through
space.  To explain this we first discuss CTAP as a precursor adiabatic passage
protocol, before
showing how this can be extended to multiple recipients.

CTAP is an all-spatial variant of the well-known STIRAP (STImulated
Raman Adiabatic Passage) technique \cite{GRSB90}.  The simplest form
of CTAP requires a single particle which can be placed in a coherent
superposition of three spatially-defined quantum states.  Canonical
CTAP effects particle motion between the two outermost sites by
adiabatic variation of the tunnel matrix elements between neighbouring
sites.  It uses the counter-intuitive pulse sequence, where the tunnel
matrix element between the sites where the particle is \emph{not}
present is initially high, whereas the tunnel matrix element connected
to the site with the particle is initially zero.  The
counter-intuitive pulse sequence affords considerable robustness to
the transport protocol and has the surprising property that the particle
is never found at the central site.  CTAP has been studied
theoretically for electrons \cite{GCHH04}, atoms \cite{ELC+04,
  EMC+06}, BECs \cite{GKW06, RCP+08}, and superconductors \cite{SBF06}, and was recently
demonstrated with photons \cite{LDVOL07}.

Techniques based on spatial adiabatic passage have a considerable
advantage over their more familiar quantum optical counter-parts.
With atomic and molecular systems, the Hilbert space of the system is
defined by the physics of the system under investigation.  By
contrast, advances in nano-fabrication give the intriguing ability to
\emph{engineer} a desired Hilbert space by technologies such as lithography or the application of spatially varying optical fields.  This ultimately provides new
flexibility and the potential to realise novel quantum devices.  CTAP
can be extended using linear schemes that extend the number of sites
over which transport can occur.  The quantum optical versions of these
extensions are the alternating \cite{SBOR91} and straddling
\cite{MT97} STIRAP schemes, and these have been investigated in the
context of CTAP.  The straddling CTAP scheme was first considered in
Ref. \cite{GCHH04} and later demonstrated with photons \cite{VOF+08},
and the alternating scheme has also been investigated \cite{PL06,
  JGC+09}.  The alternating scheme is particularly interesting for
interferometric schemes and non-trivial loop topologies because of the
transient population, which allows for sensing, and in this context a
scheme for an adiabatic electrostatic Aharanov-Bohm interferometer has
been proposed \cite{JG10}.

This paper is organised as follows.  We first discuss the extension of
the Multiple Recipient Adiabatic Passage (MRAP) protocol to account
for large scale quantum networks, and discuss some of the properties
of the null space of the resulting Hamiltonian.  We then show an
application of this protocol to the task of parallel IFM in the
quantum minefield problem.  Following this, we explore some practical
limitations in terms of ensuring the adiabaticity of the protocol.
Finally we show how this protocol can be converted into collision-free
quantum routing technique.

\section{Multiple Recipient Adiabatic Passage and quantum tree networks}

The possibility of engineering Hilbert spaces by spatial location of
quantum sites, provides many opportunities for the realisation of
novel devices.  Here we focus on the ability to engineer branched
quantum networks. One proposal for a branched adiabatic passage scheme
is Multiple Reciever Adiabatic Passage, MRAP \cite{GDH06}.  This was
investigated as a form of quantum fanout, useful for the direct
synthesis of operator measurements \cite{DGH07}.

\begin{figure}
\begin{centering}
\includegraphics[width=0.6\columnwidth]{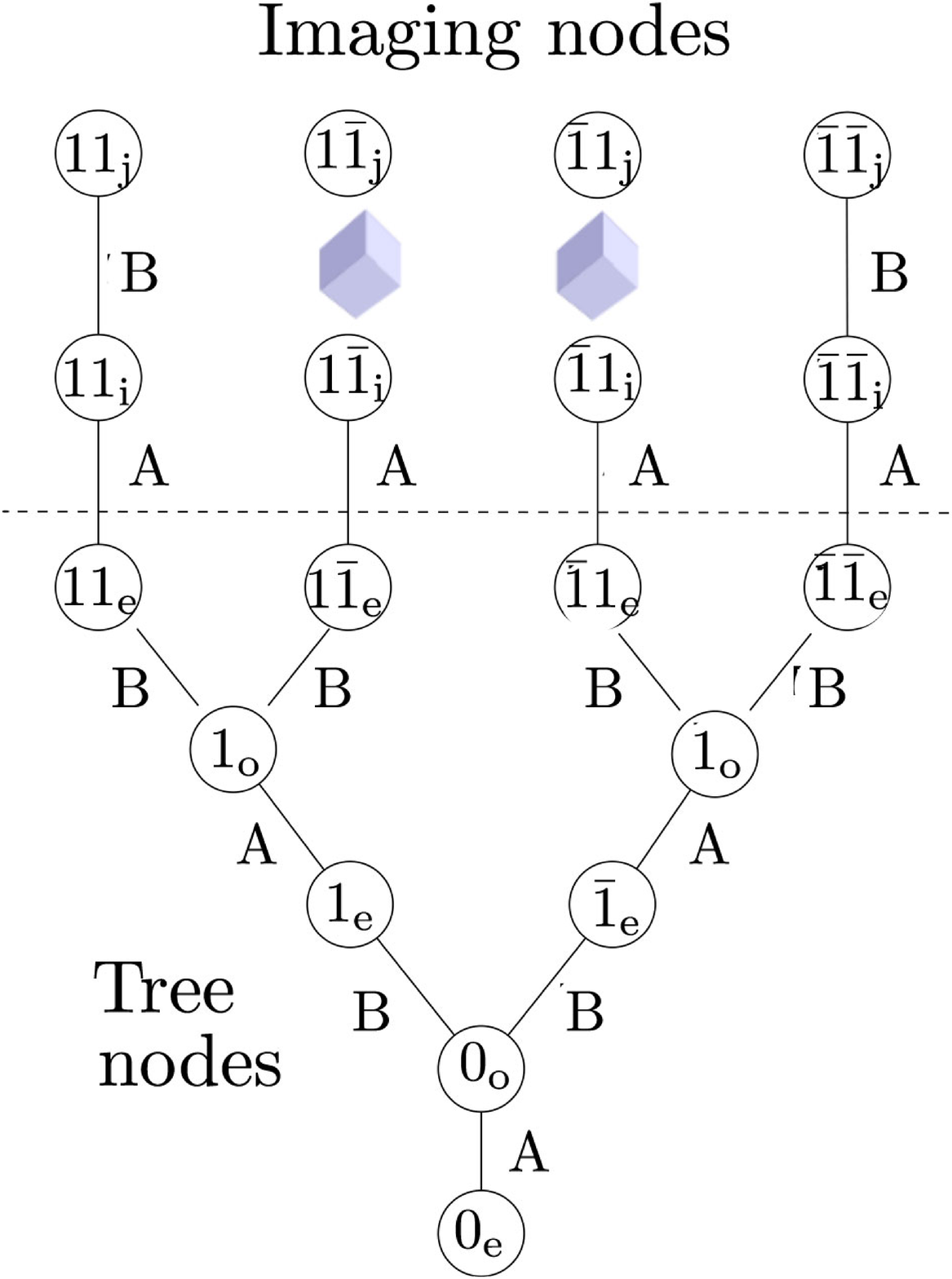}
\par\end{centering}
\caption{Schematic diagram showing the tree suitable for imaging in a
  four-leaf device. Nodes are labelled according to their position in
  the tree. Subscripts `o' and `e' represent odd and even numbers of
  edges, whilst the $i$ and $j$ nodes denote the imaging nodes.  The
  quantum minefield is realised by the quantum bombs at sites that
  break the connection between the $i$ and $j$ sites, represented by
  cubes.  In this case there are two quantum bombs, between
  $\ket{1\bar{1}_i}$ and $\ket{1\bar{1}_j}$, and $\ket{\bar{1}1_i}$
  and $\ket{\bar{1}1_j}$. The strength of interaction between two
  waveguides is labelled by `A' and `B'.\label{fig:ANDtree}}
\end{figure}

The dynamics of the MRAP tree are governed by the system Hamiltonian,
depicted schematically in Figure \ref{fig:ANDtree}.  We assume that the
energy of the particle at any site is the same, and hence the
Hamiltonian is solely defined by the adjacency matrix.  To label the
tree, first note that each node other than the zeroth node, either has
two connections or three connections, corresponding to whether there
is an odd number of links to the initial node, or an even number
respectively.  We choose to label those sites with an odd number of
links with $o$, and those with an even number $e$.  We next number all
of the $e$ sites by the path taken to reach them using a balanced
ternary notation according to Figure \ref{fig:ANDtree}, where up
pathways are denoted by adding the digit $1$ to the right hand side
(least significant digit) of the number, and down pathways by the
digit $\bar{1}$.  The $o$ sites are numerically labelled with the same
number as the site to their immediate left.

The tunnel matrix element (TME) connecting states follows the
alternating convention, which also be considered as an $A-B$ chain.
The strength of a connection between states in the order $e-o$ is $A$,
and between $o-e$ is $B$, as shown in Figure \ref{fig:ANDtree}.  The
magnitudes of all of the $A$ TMEs are the same, as are all the $B$.
The values of the couplings are varied according to the
counter-intuitive pulse sequence (also employed in CTAP). This pulse
sequence initially has coupling $B$ on, rather than the intuitive way
of operating which would raise $A$ first. For simplicity we choose a
sinusoidal variation of TME with time, $t$, i.e.
\begin{eqnarray}
A(t) = \sin^2(\pi t /2T), \quad B(t) = \cos^2(\pi t/2T),
\end{eqnarray}
where $T$ is the total time of the protocol and the maximum values of
the TMEs are normalised to unity.

The general form of our MRAP Hamiltonian is
\begin{eqnarray}
H &=& A \sum_{k} \left(a^{\dag}_{ko}a_{ke}\right) + B \sum_{k} \left(a^{\dag}_{3k+1o} a_{ke}a^{\dag}_{3k-1o} a_{ke}\right) \nonumber \\
	& &+ A \sum_{k} \left(a^{\dag}_{ki}a_{ke}\right) + B \sum_{k} \left(a^{\dag}_{kj}a_{ki}\right) + \mathrm{h.c.},
\end{eqnarray}
where $a$ ($a^{\dag}$) is the usual annihilation (creation) operator, $k$ sums over all sites in the tree.  The particle distribution via tree is achieved by the $e$ and $o$ sites, as per the first two sums in the Hamiltonian, whilst the second sums effect the CTAP like imaging plane.

The simplest form of MRAP is a four-site system, as shown in the
lowest four tree nodes of Figure \ref{fig:ANDtree}, i.e. $\ket{0_e}$,
$\ket{0_o}$, $\ket{1_e}$ and $\ket{1_o}$.  The Hamiltonian required to
understand this is equivalent to that discussed in the context of
geometric phase gates in the tripod atom \cite{UPSB04}. Our scheme
takes this structure as our starting point, and extends to realise an
MRAP tree, similar in spirit to NAND trees \cite{FGG08}.  The essence of the imaging scheme is that objects in the
imaging plane break the symmetry of the tree, and remove sections of
the Hamiltonian from the null space.  The adiabatic passage can only
explore the null space, and hence pathways that have been removed
cannot have population, thereby realising interaction-free
measurements.

Although we discuss the MRAP protocol in an arbitrary,
decoherence-free setting, it is clear that decoherence will be
deleterious to the protocol.  The ideal method of realising this
scheme would therefore be in an optical setting using waveguides such
as those demonstrated in Refs. \cite{LDVOL07,VOF+08}.  In such schemes, each site is replaced by a waveguide, and the tunnel matrix element between sites controlled by varying the proximity of the waveguides.  Varying proximity changes the evanescent tails of the modes, and hence their overlap.  In this way the \emph{temporal} variation of the adiabatic passage is translated into a \emph{spatial} evolution, but the essential characteristics of the evolution are unaltered between representations.

The Hamiltonian in matrix form of a two-leaf MRAP structure is 
\begin{eqnarray}
H_2 = \left[ 
\begin{array}{cccc}
	0 & A & 0 & 0 \\
	A & 0 & B & B \\
	0 & B & 0 & 0 \\
	0 & B & 0 & 0 \end{array} \right],
\end{eqnarray}
with basis ordering $\left\{\ket{0_e}, \ket{0_o}, \ket{1_e},
\ket{\bar{1}_e}\right\}$. The identical on-site energies have been
subtracted. The null space of this Hamiltonian is spanned by
\cite{DGH07}.
\begin{eqnarray}
\ket{D^{(2)}_1} = \frac{B\ket{0_e} - A\ket{1_e}}{\sqrt{A^2 + B^2}},\quad \ket{D^{(2)}_{\bar{1}}} = \frac{B\ket{0_e} - A\ket{\bar{1}_e}}{\sqrt{A^2 + B^2}}.
\end{eqnarray}
This spanning ensures that under adiabatic evolution, a particle
initially in $\ket{0_e}$ will be transported to a coherent superposition
$-(\ket{1_e} + \ket{\bar{1}_e})/\sqrt{2}$, without any population even
transiently being present in site $\ket{0_o}$.

To create the Hamiltonian for the entire MRAP tree, leaf nodes are
replaced by the Hamiltonian $H_2$, so for example, the four-leaf MRAP
tree would be
\begin{eqnarray}
H_4 = \left[ \begin{array}{cccccccccc}
	0 & A & 0 & 0 & 0 & 0 & 0 & 0 & 0 & 0 \\
	A & 0 & B & B & 0 & 0 & 0 & 0 & 0 & 0 \\
	0 & B & 0 & 0 & A & 0 & 0 & 0 & 0 & 0 \\
	0 & B & 0 & 0 & 0 & A & 0 & 0 & 0 & 0 \\
	0 & 0 & A & 0 & 0 & 0 & B & B & 0 & 0 \\
	0 & 0 & 0 & A & 0 & 0 & 0 & 0 & B & B \\
	0 & 0 & 0 & 0 & B & 0 & 0 & 0 & 0 & 0 \\
	0 & 0 & 0 & 0 & B & 0 & 0 & 0 & 0 & 0 \\
	0 & 0 & 0 & 0 & 0 & B & 0 & 0 & 0 & 0 \\
	0 & 0 & 0 & 0 & 0 & B & 0 & 0 & 0 & 0 \end{array} \right],
\end{eqnarray}
with basis ordering $\{\ket{0_e}$, $\ket{0_o}$, $\ket{1_e}$, $\ket{\bar{1}_e}$, $\ket{1_o}$, $\ket{\bar{1}_o}$, $\ket{11_e}$, $\ket{1\bar{1}_e}$,$\ket{\bar{1}1_e}$, $\ket{\bar{1}{1}_e}\}$.  The null space is spanned by the vectors
\begin{eqnarray}
\ket{D_{11}^{(3)}} &=& \frac{B^2\ket{0_e} - AB \ket{1_e} + A^2\ket{11_e}}{\sqrt{A^4 + A^2B^2 + B^4}}, \\
\ket{D_{1\bar{1}}^{(3)}} &=& \frac{B^2\ket{0_e} - AB \ket{1_e} + A^2\ket{1\bar{1}_e}}{\sqrt{A^4 + A^2B^2 + B^4}}, \\
\ket{D_{\bar{1}1}^{(3)}} &=& \frac{B^2\ket{0_e} - AB \ket{\bar{1}_e} + A^2\ket{\bar{1}1_e}}{\sqrt{A^4 + A^2B^2 + B^4}}, \\
\ket{D_{\bar{1}\bar{1}}^{(3)}} &=& \frac{B^2\ket{0_e} - AB \ket{\bar{1}_e} + A^2\ket{\bar{1}\bar{1}_e}}{\sqrt{A^4 + A^2B^2 + B^4}}, 
\end{eqnarray}
where the superscript denotes the number of non-zero positional states contributing to the null state, and the subscript the numerical value of the final (leaf) state.  
%\begin{widetext}
Any state which is a superposition of the null states is also in the null space.  So therefore the state which transfers a particle from the base to the equally weighted superposition of leaf nodes is the normalised sum of these states, i.e.
\begin{eqnarray}
\frac{2B^2\ket{0_e} - AB \left(\ket{1_e} + \ket{\bar{1}_e}\right) + \frac{A^2}{2} \left(\ket{11_e} + \ket{1\bar{1}_e} + \ket{\bar{1}1_e} + \ket{\bar{1}\bar{1}_e}\right)}{\sqrt{A^4 + 2A^2B^2 + 4B^4}}. %\nonumber \\
\end{eqnarray}
%\end{widetext}

Our notation allows us inductively explore both the total tree
structure, and the vectors spanning the null space.  So for example,
consider adding another MRAP-type branch to an even node connected to site
$\ket{0_e}$ via a null state $\ket{D_i^{(j)}}$. The effect of adding
these new links is to replace the state $\ket{D_i^{(j)}}$ with two new
states, which are
\begin{eqnarray}
\ket{D_{3i+1}^{(j+1)}} &=& B \ket{D_{i}^{(j)}} + A^{j} \ket{3i+1}, \\
\ket{D_{3i-1}^{(j+1)}} &=& B \ket{D_{i}^{(j)}} + A^{j} \ket{3i-1},
\end{eqnarray}
where the subscript on the $D$ is represented in decimal form for simplicity.  Although the tree structures are shown as if all leaf nodes are the same distance from the origin, in fact there is no requirement for this and the essential features of the transport protocol are unaffected by leaf nodes of varying distance to the origin.  

With regard to interaction free measurement, the inductive method for
generating the null states is particularly useful for understanding
the properties of the network as a whole, but we can also use this
method essentially \emph{in reverse} to understand the effect of
removing couplings from the network.  Observe that for every null
state, the population in all of the $o$ nodes is zero. Trivially, if
the TME connecting to a site with zero population on the left is also
zero, then the population in any subsequent sites must be zero.  In
other words, a break in the chain immediately after an $o$ site
ensures that there cannot be any population in the remaining sites to
the right of the break.  Their contributions are removed from the null
space.  This insight immediately allows us to base an interaction free
measurement protocol based on MRAP networks.

\section{MRAP for imaging the quantum minefield}

Here we turn to the task of how to use our protocol on a parallel
IFM of a quantum minefield.  This is a situation where there are a
number of sites $n$, and a number of quantum bombs $m<n$, where $m$
may be unknown.  Our task is to determine the locations of the bombs
without setting them off.  This task may be accomplished by the
network shown in Figure~\ref{fig:ANDtree}, where the minefield is
located between the sites labelled $i$ and $j$.  Because breaks in the
coupling for the tree remove states from contributing to the null
space, and because the population in the site immediately before the
broken chain is zero, the adiabaticity of the transport protocol
forces every photon to avoid the broken link.  This satisfies all of
the requirements of an IFM.  To more explicitly show how MRAP can be
used for imaging objects, Figure \ref{fig:ANDtree} shows a typical
configuration where this can be achieved. An excitation, such as a
photon, starting from the root of the tree, $\ket{0_e}$, is
transported via adiabatic passage to the leaves of the tree. An
object, placed just before the leaf nodes, blocks the TME on the final
$o-e$ link (or not) revealing its presence or absence. This happens
despite the fact that the photon is never found in a location adjacent
to the bomb.

The observation that the null space of the MRAP tree is spanned by all
of the null vectors that connect the base to the leaves immediately
suggests a parallel IFM scheme.  By altering the symmetry of a
pathway, we can remove it from the null space, \emph{without affecting
  the other null vectors}. Consider the tree depicted in
Figure~\ref{fig:ANDtree}.  In this arrangement, the MRAP tree is
terminated by three-site CTAP pathways, which preserve the overall
symmetry of the null space.  However by removing the TME between the
final two sites, the symmetry of the pathway is destroyed, and the
vector connecting the leaf to the base of the tree is removed from the
null space.  We term the line between the final two sites as the
imaging plane, and the object in this case is assumed to occlude some
subset of the paths.  A particle adiabatically evolving through the
network will be unable to explore the occluded pathways, and neither
will it be able to occupy the site directly before the object.  Hence
the particle performs an IFM of the image plane in parallel, with the
fidelity set by the adiabaticity of the overall protocol.

%The MRAP protocol for interaction free measurement consists of two
%interconnected parts.  First the initial state is fanned out using a
%tree structure, labeled \emph{tree nodes} in Figure
%\ref{fig:ANDtree}. In this figure a single level of the tree is shown,
%however the tree can be branched out to many more levels. After these
%nodes, the excitation enters pairs of \emph{imaging nodes}. These
%nodes are either coupled (in the absence of anything to image) or the
%uncoupled (because of an opaque object between the two locations). The
%aim of interaction free imaging is to obtain an image of the object,
%in this case residing between nodes $\ket 5$ and $\ket 7$ without ever
%significantly populating either of these two states.

In the arrangement shown in Figure \ref{fig:ANDtree}, the population
of all odd states (eg. $\ket{0_o}$ and $\ket{1_o}$), in addition to the
states $\ket{1\bar{1}_e}$ and $\ket{\bar{1}1_e}$ is (in the adiabatic
limit) always zero. This follows from the fact that that neighbours of
leaf nodes always have zero amplitude. Therefore, any exitation in the
system has no probability of actually interacting with the object
being imaged (or even undergoing adiabatic passage down a branch of
the tree in which the an object is located). In this example, the
excitation undergoes adiabatic passage from state $\ket{0}$ to states
$\ket{11_j}$ and $\ket{\bar{1}\bar{1}_j}$ where it can be
safely detected. Similarly for any combination of objects obscuring the imaging
nodes, the excitation proceeds directly to the imaging nodes which are
not occulded, and never undergoes adiabatic passage down the branches
of the tree which lead to an object.

Returning to the initial quantum bomb analogy, we can now imagine the
parallel version - a quantum minefield placed in the imaging plane.
The sensing particles will adiabatically explore the minefield, and
will be unable to interact with sites with bombs (as they are removed
from the null space).  Measurements of the particles at the leaf nodes
will definitively mark sites without bombs, and eventually the pattern
of the bombs will be discovered to arbitrarily high accuracy, without
the loss of \emph{any} particles.

\begin{figure}
\begin{centering}
\includegraphics[width=8cm]{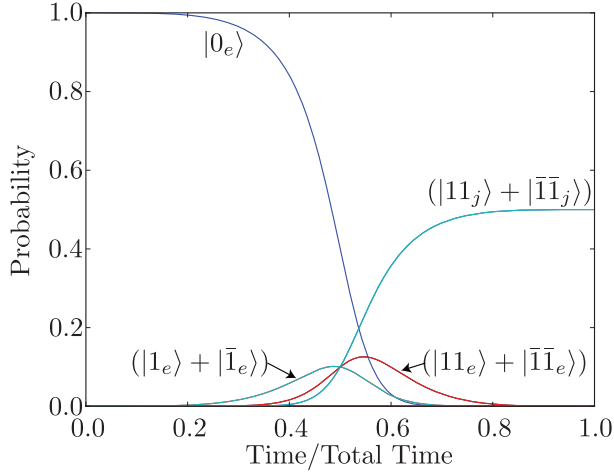}
\par\end{centering}
\caption{Adiabatic passage showing the evolution of the wave-function
  for the tree shown in Figure~\ref{fig:ANDtree}, as a function of the fractional time through the protocol, with two obscured imaging
  nodes ($\ket{1\bar{1}_{j}}$ and $\ket{\bar{1}1_{j}}$).  The initial population is all in the state $\ket{0_{e}}$ and smoothly evolves to the superposition $(\ket{11_{j}} + \ket{\bar{1}\bar{1}_{j}})/\sqrt{2}$, with transient evolution in the states $\ket{1_{e}}$, $\ket{\bar{1}_{e}}$, $\ket{11_{e}}$ and $\ket{\bar{1}\bar{1}_{e}}$. \label{fig:ANDgraph}}
\end{figure}

A typical evolution is shown in Figure \ref{fig:ANDgraph}.
%, with accompanying animation depicted in Figure~\ref{fig:animation}. 
Here we show the results of evolution for the case depicted in Figure~\ref{fig:ANDtree}, i.e. with two obscured imaging sites ($\ket{1\bar{1}_{\mathrm{j}}}$ and $\ket{\bar{1}1_{\mathrm{j}}}$).  If we write the state as $\ket{\psi}$, then the evolution is represented by $|\langle i | \psi \rangle|^2$ for all sites $\ket{i}$. In this
case adiabatic passage takes the excitation from the initial state
$\ket{0_\mathrm{e}}$, to the superposition $(\ket{11_{\mathrm{j}}} + \ket{\bar{1}\bar{1}_{\mathrm{j}}})/\sqrt{2}$.  In the adiabatic limit, all of the $\mathrm{o}$ and $\mathrm{j}$ sites are unpopulated, as are $\ket{1\bar{1}_{\mathrm{e}}}$,   $\ket{1\bar{1}_{\mathrm{j}}}$, $\ket{\bar{1}1_{\mathrm{e}}}$, and $\ket{\bar{1}1_{\mathrm{j}}}$, as expected by the IFM. Transient population is observed in intermediate states, as expected for the alternating protocol.  The first maximum corresponds to the superposition (at time $t \approx 0.49$) $\ket{1_{\mathrm{e}}} + \ket{\bar{1}_{\mathrm{e}}}$, whilst the second maximum (at time $t \approx 0.55$) to the superposition $\ket{11_{\mathrm{e}}} + \ket{\bar{1}\bar{1}_{\mathrm{e}}}$

\begin{figure}
\begin{centering}
\includegraphics[width=8cm]{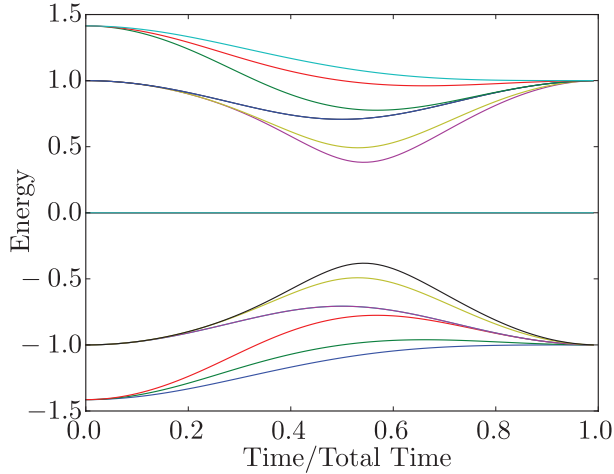}
\par\end{centering}
\caption{Energy level diagram for a two-level tree, with no object obscuring the imaging plane. The minimum energy gap occurs approximately halfway through the protocol, in accordance with expectations from other adiabatic passage protocols (CTAP, STIRAP and MRAP).  Robustness of the protocol is guaranteed provided the rate of evolution is slow compared with this minimum energy gap.  \label{fig:energyLevel}}
\end{figure}

%\emph{Energy Gap}: 
To ensure adiabatic evolution, a
critical consideration is the energy gap between the ground ($E=0$)
state and the first excited state. A typical energy level diagram for
a two-level tree, with no object obscuring the imaging plane, is shown in Figure \ref{fig:energyLevel}.

Initially, when $A=0$, $B=1$ there are five distict, but degenerate
energy levels. These are at $E=0$, which has been discussed in detail
in this paper, at $E=\pm 1$ which are contributions from symmetric and
anti-symmetric superpositions the imaging plane, and at $E=\pm
\sqrt{2}$ corresponding to symmetric and anti-symmetric superpositions
across different levels of the tree.

\begin{figure}
\begin{centering}
\includegraphics[width=8cm]{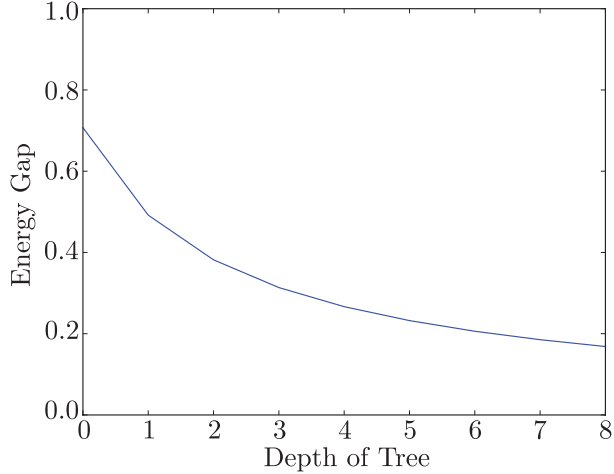}
\par\end{centering}
\caption{Minimum energy gap between the $E=0$ state, and the first
  excited state for different sized trees, for the case that $A_{\max} = B_{\max} = 1$.  Note the monotonic decrease in the energy gap with increasing tree depth, which implies that evolution through more complicated trees must proceed commensurately more slowly than for shorter depth trees. \label{fig:gap}}
\end{figure}

As evolution proceeds, the minimum energy gap between the $E=0$ state
and the excited states narrows. According to the adiabaticity criteria, for any two eigenstates $\ket{\psi}$ and $\ket{\phi}$, adiabatic evolution is assured when
\begin{equation}
\frac{|\bra{\psi} \dot{H} \ket{\phi}|}{\left(E_{\ket{\psi}} -  E_{\ket{\phi}}\right)^2} \ll 1.
\end{equation}
For this reason, the energy gap is of crucial importance for maintaining the adiabaticity of
the operation. Fortunately, for any reasonable level of tree, it is
possible to determine the size of this gap, and show that it does not
grow too small. Figure \ref{fig:gap} shows a numerical calculation of
the size of the minimum energy gap for up to a depth eight,
or $256$ imaging nodes.

%\emph{Extending the MRAP Tree}: 
This example may be extended to much
larger trees, of much larger depth is a straightforward
way. Additional nodes may be added in a tree-like structure to
provide many more leaf nodes, and therefore many more locations for
interaction free imaging.

If objects are placed in the imaging plane, then it is possible
(after several iterations of the MRAP procedure) to determine the
position of these objects to arbitrary accuracy.  It should be noted that if objects can be
placed at any level of the tree, it is only possible to determine
which branches of the tree have been obscured by the objects, but not
necessarily to determine their exact positions.  For example, in our scheme one cannot distinguish between two bombs placed immediately before sites $\ket{11_j}$ and $\ket{1\bar{1}_j}$ and one bomb placed immediately before site $\ket{1_o}$.  Care should also be taken to ensure that not \emph{all} paths to
imaging nodes are obscured. If this is the case, then when $A\neq 0$,
$B\neq 0$ there is no longer a valid $E=0$ state, and it is possible
to observe beating in the system.

\section{Collision-free routing}

In this section we point out that the MRAP scheme as discussed offers another surprising result arising from the fact that a \emph{local} change in the Hamiltonian produces a \emph{global} alteration of the null states.  This feature gives rise to the possibility of collision-free routing, which we describe below.

Imagine a distribution network where particles (possibly containing information, or halves of Bell pairs) need to be distributed from a starting node, to multiple recipients.  Examples of protocols that could take advantage of such a network are quantum cryptographic networks where a base station wants to share random numbers for use as quantum keys with many users.  So for concreteness let us assume that we are employing the MRAP tree to distribute qubits, and that there is no coupling between the qubit degree of freedom and the spatial degree of freedom.  Returning to Figure \ref{fig:ANDtree} imagine that Alice is at site $\ket{0}$, and wants to distribute qubits to Bobs at the leaf nodes of the MRAP tree.  Alice doesn't care which qubit goes to which Bob, but she does need to know the time stamp of arrival of the qubits (so that appropriate correlations can be made during public basis comparison).  Let us further assume that the Bobs need a certain amount of time to process the arrival of a particle (detector dead time), or that for other reasons (for example they have a full buffer of qubits) they do not wish to receive qubits.  However, they do not wish to communicate with Alice or the Bobs when they want to receive particles.  

In a classical setting with conventional routing, Alice would need to decide to which Bob she would send a particle.  To avoid the dead time problem, there are many solutions which involve some scheduling of the distribution of qubits, however it seems that any such system will either enforce an effective clock rate, and possibly entail the `collision' of qubits in the network, for example if two qubits are sent to the same receiver (Bob) during the dead time of the detection network.  Remarkably, the MRAP approach provides an elegant solution to this problem.

For the IFM protocol discussed above, the breaking of a connection led
to an alteration of the symmetry of the null space, thereby removing
certain states from the null space.  Adiabaticity of evolution then
prevented particles from exploring the removed states.  However,
another way to remove states from the null space is simply to apply a
small energy shift to the state.  For example, if one of the Bobs, say
at site $\ket{11}$ in Figure \ref{fig:ANDtree} shifts the energy of
their site, they will immediately remove their site from the null
space, without affecting the rest of the null space.  This result is
interesting.  It means that a local change in the Hamiltonian, affects
the entire null space.  However this change is not observable by the
other Bobs or Alice.  The net result is that to have a probability of
receiving particles, the Bobs should ensure that their receiving site
has zero energy.  But if it any time they wish to cease receiving
qubits, then they shift their energy.  This could be because of
detector dead time, or simply because they do not require qubits.
Each Bob will receive particles randomly, collecting a fraction of the
distributed particles sent by Alice equal to the rate of particles
transmitted divided by the number of Bobs receiving particles.
Finally, Alice need not know who is on the network, she only needs to
continue transmitting at the rate available to her.

This surprising routing solution is interesting, and would seem to be
ideally applied in an optical setting, again taking advantage of the
structures similar to the CTAP waveguides networks demonstrated by
Longhi and co-workers \cite{LDVOL07, VOF+08}.

\section{Conclusion} 

We have explored the extension of the MRAP (Multiple Recipient
Adiabatic Passage) procotol to explore adiabatic passage through
tree-like structures.  This MRAP tree affords a new method to realise
interaction free measurement (IFM).  The original quantum bomb problem was non-deterministic, and could only sense a single bomb.  Our approach 
allows for deterministic parallel IFM  arising from
the adiabaticity of the protocol, and the perturbation to the symmetry
of the null space introduced by the objects being imaged.  Due the adiabaticity of the protocol, this approach should be robust with respect to perturbations in the distribution network.

We have
also shown that with a minor variation to the protocol, it can be used
to effect a novel distribution network, where collision-free routing
can be achieved.  In particular where source distributes particles, but
does not need to specify any of the routing.  We are not aware of any
classical protocol that can achieve these outcomes.  

One issue with
the scheme is that the MRAP protocol requires spatial coherence of the
particle exploring the network.  For this reason, a photonic
implementation would seem to be the only practical system to realise
this network, although it is clear from the quantum mechanics that in
principle, any quantum system could exploit these effects providing
that the spatial decoherence rate is sufficiently long.

\ack The authors would like to thank Andy Martin for useful
discussions.  This project is supported by the Australian Research
Council Centre of Excellence Scheme (CE110001027). L.H. is supported
under the Australian Professorial Fellowship scheme
(DP0770715). A.D.G. acknowledges the Australian Research Council for
financial support (Project No. DP0880466).

\section*{References}

\bibliographystyle{iopart-num}
\bibliography{papers}

\end{document}